\documentclass[12pt]{article}
\usepackage{amsmath, amsfonts, bbm}
\usepackage{graphicx,psfrag,epsf}
\usepackage{enumerate}
\usepackage[round]{natbib}
\usepackage[titletoc,title]{appendix}
\usepackage{url} 

\newcommand{\blind}{0}

\usepackage{color} 

\usepackage{soul}

\newcommand{\gjh}[1]{\textcolor{black}{#1}}
\newcommand{\strike}[1]{}
\newcommand{\cb}[1]{\textcolor{black}{#1}}

\begin{document}

\def\spacingset#1{\renewcommand{\baselinestretch}%
{#1}\small\normalsize} \spacingset{1}

\if0\blind
{
  \title{\bf An Expectation Maximization Algorithm for High-Dimensional Model Selection for the Ising Model with Misclassified States}
  \author{David \cb{G.} Sinclair and Giles Hooker \thanks{
    \cb{David} Sinclair is PhD Candidate, Department of Statistical Science, Cornell University, 301 Malott Hall, Ithaca, NY 14853 (Email: dgs242@cornell.edu). \cb{Giles }\gjh{Hooker is Associate Professor of Biological Statistics and Computational Biology, Cornell University, 1186 Comstock Hall, Ithaca, NY 14853 (Email: gjh27@cornell.edu). The authors gratefully acknowledge support from grants NSF DMS-1053252 and NSF DEB-1353039}}}
    \date{}
  \maketitle
} \fi

\if1\blind
{
  \bigskip
  \bigskip
  \bigskip
  \begin{center}
    {\LARGE\bf An Expectation Maximization Algorithm for High-Dimensional Model Selection for the Ising Model with Misclassified States}
\end{center}
  \medskip
} \fi

\bigskip
\begin{abstract}
We propose the misclassified Ising Model; a framework for analyzing dependent binary data where the binary state is susceptible to error.  We extend the theoretical results of the model selection method presented in  \citet{ravikumar2010high} to show that the method will still correctly identify edges in the underlying graphical model under suitable misclassification settings.  With knowledge of the misclassification process, an expectation maximization algorithm is developed that accounts for misclassification during model selection.   We illustrate the increase of performance of the proposed expectation maximization algorithm with simulated data, and using data from a functional magnetic resonance imaging analysis.
\end{abstract}

\noindent%
\vfill

\newpage
\spacingset{1.45} 

\section{Introduction}

This paper proposes an extension of estimation methods for graphical models to cases where node values are observed with error. In particular, motivated by data from functional magnetic resonance imaging (fMRI), we examine the consequences of misclassification noise in an Ising network model on estimation methods proposed in \citet{ravikumar2010high} and show that the estimated edge set can be improved by accounting for misclassification rates.

Graphical models have proven to be a useful tool in modeling a wide range of data, arising in fields such as neuroscience, genetics, social networks, image restoration, traffic models, and disease case modeling, among many.   The graph structure provides a useful mathematical framework for representing  complex dependencies among a large collection of objects.

In this paper we focus on undirected graphical models, which are specified by a graph $\mathcal{G} = (V, E)$ for a node set $V = \{1, 2, \dots, p\}$ and an edge set $E \subset V \times V$.  A random vector with this graph structure is assumed to follow the Markov Property \citep{kindermann1980markov}\gjh{: \strike{.  A consequence of the pariwise Markov Property is that}} the $i^{th}$ and $j^{th}$ element of the vector are dependent conditional on the remaining nodes if and only if $(i, j) \in E$.  Thus, we are concerned with uncovering the structure of the edge set $E$ and therefore uncover\cb{ing} conditional dependencies within our dataset.

Further, we assume that our data is binary where the dependencies are entirely captured by pariwise relationships \gjh{resulting in \strike{.  T}t}he Ising Model \citep{ising1925beitrag}, \gjh{\strike{which will be}} detailed in Section \ref{istheory}, \cb{which} corresponds precisely to these assumptions.  The Ising Model has proven useful in data analysis settings such as functional magnetic resonance imaging (fMRI) \citep{sinclair2017}, image restoration \citep{kandes2008statistical,geman1984stochastic}, spatial statistics \citep{banerjee2014hierarchical}, social network analysis \citep{montanari2010spread}, and genetics \citep{majewski2001ising}.

Structure learning of the edge set in the Ising model is a well-studied problem in the statistics literature.   Considerable attention has been given to finding theoretic information bounds for learning Ising graph structures \citep{scarlett2016difficulty,tandon2014information,santhanam2012information}.  Table 1 in \citet{scarlett2016difficulty} gives a useful summary of the graphical assumptions for which these information theoretic bounds are known.

Due to the computational intractability  of the partition function $Z(\theta^*)$ for the Ising distribution function given in Equation (\ref{eq:1}) \citep[see][]{welsh1993complexity}, various approaches have been developed in order to perform sound statistical methodology under this practical constraint.

\citet{barber2015high} show an extended BIC method for uncovering the underlying graph in the Ising data setting with theoretical bounds.  \citet{bresler2015efficiently} develop a greedy algorithm, which uses a structural property of mutual information associated with Ising models to prove asymptotic exact learning of the underlying graph.  \citet{ravikumar2010high} show theoretic bounds for a neighborhood-based regularized logistic regression approach for performing model selection  analogous to the Meinshausen-B{\"u}hlmann approach for Gaussian graphical models \citep{meinshausen2006high}.

One potential issue with categorical data is the possibility for misclassification. \gjh{This arises in fMRI data where \strike{the activation state in each component of a parcellation of the brain is obtained by a threshhold on the BOLD response.}} \cb{the traditional General Linear Model approach attempts to find areas of the cortex that have been significantly activated, which corresponds to a threshold of the BOLD response's association with the HRF function} \cb{\citep{lindquist2008statistical}.  When the cortex is reduced to specialized regions via a parcellation \citep{sinclair2017, gordon2016generation} w}\gjh{\strike{W}e can think of this procedure as assigning a latent label to each parcel and may suspect possible misclassification \cb{\strike{where the latent label} when the BOLD respone's association with the HRF} is close to the threshold. }  If there is a non-zero probability of misclassification,  it can be shown that the data no longer follows an Ising distribution, and thus it is not clear if current structure learning methods can still perform adequately.

In this paper we extend theory behind \citet{ravikumar2010high}'s approach to handle misclassification and, conditional on this result, we develop a methodology for further boosting of structural learning performance via an {\it expectation maximization} (EM) technique \citep{dempster1977maximum} that can be used if there is knowledge of the misclassification process.  Due to the inherent dependency in our data set, it is difficult to show that the EM method will always increase the marginal log likelihood\gjh{. H}owever we show that if the learned structural dependency can predict a candidate state with high probability, the EM method can provide gains in efficiency.

In Section \ref{istheory} \gjh{of this paper} the misclassified Ising model is defined, and theoretical guarantees are stated.  In Section \ref{EM} the algorithm for incorporating misclassification information in an updated edge set estimated is described.  Section \ref{sims} looks at simulations to better understand the performance of this methodology.  Section \ref{fmri} shows how this methodology can be applied in an fMRI setting, and simulations are done to show the method should still increase structural learning accuracy.

\section{Misclassified Ising Model and Theoretical Guarantees}\label{istheory}

In this section we develop the Misclassified Ising Model, and discuss theoretical guarantees for estimating the underlying edge set with this added noise assumption.

\subsection{Ising Model}\label{istheory:1}
We focus on the special case of the Ising Model as described in \citep{ravikumar2010high}, which we refer to as the $Ising(G, \theta^*)$ distribution.  Let $\mathfrak{X} = (x^{(1)}, \dots, x^{(n)})$ be $n$ i.i.d. observations of $X = (x_1, \dots,x_p)  \sim Ising(G, \theta^*)$  in which $x_s \in \{-1, 1\}$, and $\theta^*_{st} \in \mathbb{R}$ for each $s \in V$, with probability mass function
\begin{equation} \label{eq:1}
P_{\theta^*}(x) = \frac{1}{Z(\theta^*)}\exp\left\{\sum_{(s, t)\in E}\theta_{st}^*x_sx_t\right\}
\end{equation}

Here the partition function $Z(\theta^*)$ ensures the distribution sums to one.  Recall that $\theta^*_{st} \neq 0 \iff (s, t) \in E$, and therefore our goal is to determine the support of $\theta^*$.

Due to the computational intractability of the partition function \citep{welsh1993complexity}, a neighborhood based likelihood method is adopted in \citep{ravikumar2010high}, a technique akin to the \citet{meinshausen2006high} method for Gaussian graphical models \citep{lauritzen1996graphical}, where a model selection is undertaken to find the neighborhood of each node separately.  The estimated edge set is then consolidated from the neighborhood sets.

\subsection{$\ell_1$-regularized Neighborhood-based Model Selection}\label{istheory:2}

The Ising Model has the useful property that the conditional distribution of a node takes the form of a logistic regression with the canonical link function on all remaining nodes.  Therefore, if we let $\theta^*_{\setminus r} = \{\theta^*_{ru} ; u \in V \setminus \{r\}\}$ be the edge weights associated with the node $r$, a model selection can be done via an $\ell_1$-regularized logistic regression on each node $r$ \citep{glmnet}\gjh{:}
\begin{equation}\label{neilas}
\hat{\theta}_{\setminus r} = \arg\min_{\theta_{\setminus r}\in \mathbb{R}^{p-1}} \left\{-\frac{1}{n}\sum_{i=1}^n \log P_{\theta_{\setminus r}}(x_r^{(i)} | x_{\setminus r}^{(i)}) + \lambda_{n, d, p}\|\theta_{\setminus r}\|_1\right\}
\end{equation}

In this equation, $d$ is the maximal neighborhood size, and $P_{\theta}$ is the logistic regression function with a canonical link with response $\mathbf{1}(x_{r}^{(i)} = 1)$, regression parameters $2\theta_{\setminus r}$, and predictors $x_{\setminus r} = \{x_t | t \in V \setminus \{r\}\}$.  Doing this regularized regression over each node can give us an estimate for the edge set $E$  as follows\gjh{:}
\begin{equation}\label{eq:3}
\hat{E}_{\ell_1} = \{(s, t) ; (\hat{\theta}_{\setminus s})_t \neq 0 \text{ and } (\hat\theta_{\setminus t})_s \neq 0\}
\end{equation}

In this formulation of the estimated edge set, an edge will be selected between two nodes if the corresponding estimated neighborhood sets \cb{both} contain these two nodes.

This method is shown in \citet{ravikumar2010high} to give a consistent estimate $\hat{E}_{\ell_1}$ in the sense that $P(\hat{E}_{\ell_1} = E) \rightarrow 1$ as $n \rightarrow \infty$, when $n = \Omega(d^3 \log p)$ for appropriately chosen $\lambda_{n,d,p}$.  We refer to the method for obtaining \gjh{this edge set} as {\it RWL} \gjh{in recognition of its authors}.

\subsection{Misclassified Ising Model}\label{istheory:3}

Here we introduce a formalization of the Misclassified Ising Model, which will be defined hierarchically.

We continue to assume $X \sim Ising(G, \theta^*)$, but define $\tilde{X}$ as the random vector such that $P(\tilde{X} \equiv Y| X) = \prod_{s \in V} P(\tilde{x}_s = y_s | x_s) = \prod_{s \in V} (\gamma_s^{\mathbbm{1}(y_s \neq x_s)}(1-\gamma_s)^{\mathbbm{1}(y_s = x_s) })$ for all $Y \in \{-1, 1\}^p$.  In this sense, each node is misclassified with some probability $\gamma_s$ and the misclassification is independent across nodes.  As we only observe the misclassified nodes, $\tilde{X}$, we define their distribution unconditional of $X$ as the {\it Misclassified Ising Model}, $\tilde{X} \sim MIsing_{\gamma}(G, \theta^*)$.  The theoretical guarantees for {\it RWL} under this distribution shown in Section \ref{istheory:4} do not directly assume independence of the misclassification probabilities, however this assumption is used when completing the EM update algorithm in Section \ref{EM}.

As with the Ising Model, let $\tilde{\mathfrak{X}} = (\tilde{x}^{(1)}, \dots, \tilde{x}^{(n)})$ be n i.i.d. observations of $\tilde{X}$.

\subsection{Theoretical Guarantees}\label{istheory:4}

In this section we show that when the extra noise due to misclassification is small, the estimated edge set $\hat{E}_{\ell_1}$ can still produce a reasonable model selection method.  The amount that the added noise hinders our ability to detect edges is captured by the expectation of the score function for each node-conditional distribution for the (not misclassified) Ising Model, where expectation is calculated over the true misclassified Ising Model.  Indeed, as misclassification goes to 0, the expectation of the score function goes to 0, which implies that we there is no hinderance in obtaining the edges, as expected.

Formally, $W^n_r(\theta) = -\nabla \log P_{\theta_{\setminus r}}(\tilde{x}_r^{(i)}|\tilde{x}^{(i)}_{\setminus r})$ is the score function for $P_{\theta_{\setminus r}}$ defined in equation (\ref{neilas}).  We define the {\it misclassified score} and {\it misclassified information} as
\begin{align}
S_{max} &= \max_{r \in V} |E(W^n_r(\theta^*))|\\
\tilde{Q}^*_r &=  -E(\nabla W_r^n(\theta^*))
\end{align}

Note that both of these expectations are over the misclassified distribution.  The {\it misclassified score} $S_{max}$ corresponds to the largest deviation of the expected score function over the misclassified distribution from 0.

The first two assumptions we make for our extension, are very similar to those given in \citet{ravikumar2010high}, however they are made on the {\it misclassified information} matrix.  These are stated explicitly in Appendix A, and are referred to as ($\tilde{A1}$) and ($\tilde{A2}$).    The third assumption is stated here as:

{\it ($\tilde{A3}$) Misclassification Condition}. For $C_{min}$ and $D_{max}$ as defined in ($\tilde{A1}$), and $\alpha$ as defined in ($\tilde{A2}$), we assume
\begin{equation}
S_{max} \leq \frac{C^2_{min}\alpha^2}{400D_{max}d(2-\alpha)^2}
\end{equation}

If we make the same population assumptions as given in \citet{ravikumar2010high} on the underlying Ising Model (stated in Appendix A.1), then for $\alpha$ satisfying ($\tilde{A2}$) we have the following result that corresponds to Theorem 1 in \citet{ravikumar2010high}.

\noindent\textbf{Extended Theorem 1:}  \emph{Consider an Misclassified Ising graphical model, $MIsing_\gamma(G, \theta^*)$ with parameter vector $\theta^*$ and associated edge set $E^*$ such that conditions $(\tilde{A}1)$ and $(\tilde{A}2)$ are satisfied by the misclassified information matrix $\tilde{Q}^*_r$ for all $r \in V$.  Assume the misclassified score, $S_{max}$ satisfies $(\tilde{A}3)$ and let $\tilde{\mathfrak{X}}$ be a set of n i.i.d. samples for the misclassified Ising model.  Suppose that the regularization parameter $\lambda_n$ is selected to satisfy \begin{equation} \lambda_n \geq \frac{16(2-\alpha)}{\alpha}\left(\sqrt{\frac{\log p}{n}} + \frac{S_{max}}{4}\right)\end{equation} Then there exists positive constants L and K, independent of (n,d,p) such that if \begin{equation} n > Ld^3\log p\end{equation} then the following properties hold with probability at least $1-2\exp(-K\tilde\lambda^2_nn)$, where $\tilde{\lambda_n} = \lambda_n - \frac{4(2-\alpha)}{\alpha}S_{max}$.
\begin{itemize}
\item[(a)] For each node $r \in V$ the $\ell_1$-regularized logistic regression has a unique solution and therefore uniquely specifies a neighborhood $\hat{N}(r)$.
\item[(b)] For each node $r \in V$ the the estimated neighborhood $\hat{N}(r)$ correctly excludes all edges not in the true neighborhood. Moreover, it correctly includes all
edges (r,t) for which $|\theta^*_{rt}|\geq \frac{10}{C_{min}} \sqrt{d}\lambda_n$. \label{partb}
\end{itemize}}

The proof of this result is located in Appendix A.

An interesting consequence from this result is that as $n \rightarrow \infty$ the tuning parameter does not go to 0, unless $S_{max}$ also goes to 0.  This means that by part (b) some \cb{edges} may never be correctly included with high probability due the conditional independencies of the graphical model being overcome by the misclassification.

\section{EM Algorithm for Updating Edges of $\hat{E}_{\ell_1}$}\label{EM}

\gjh{ \strike{???? This appears different from the theoretical framework above: we only assume a subset of nodes to be potentially misclassified. If so, we should note the distinction.}}

\cb{We develop an EM algorithm for obtaining an updated edge set.  In Section \ref{istheory:3}, all nodes could potentially have some amount of misclassification probability, however throughout the use of this update we assume that only a subset of nodes can be misclassified.  The distinction does not affect the related proofs for the method, although for the method to be computationally tractable the number of potentially misclassified nodes must be relatively small.}

Conditional on the initial {\it RWL} fit, \cb{\strike{giving us} resulting in} edge set $\hat{E}_{\ell_1}$ and parameter $\hat\theta_{\setminus r}$, we develop an EM-type algorithm for updating the neighborhood for certain nodes in our graphical model.  The method is run on each node individually similar to {\it RWL}.  In the usual EM approach the average joint log likelihood of the observed and latent variables is maximized in order to increase the likelihood marginally on the observed data.  Due to the complexity of the distribution in the joint case, it is difficult to maximize the log likelihood over all possible latent states.

We instead show in Appendix \ref{apEM} that maximizing the conditional distributions will still serve to increase the marginal likelihood given that the probability that a node is in the incorrect state is close to 1.  By leveraging dependency information from the initial {\it RWL} fit, we show in simulations that this condition is satisfied and we are able to increase the marginal likelihood.

In doing our EM update we focus on neighborhoods surrounding nodes that have potentially been misclassified.  In order to do this we assume we have some knowledge of the probability of misclassification for each node.  This probability can be an average misclassification over all observations for a given node, although the model has better performance when misclassified probabilities are known for each observation \cb{\strike{as well}}. Misclassification probabilities can be estimated \cb{within} each observation \cb{across nodes} if, for example, a separate EM algorithm is used to determine the state of each node, then the latent variable state probabilities correspond to the probability of misclassification.  In \citet{sinclair2017}, misclassification probabilities can be derived from the implicit mixture model for continuous signaling in fMRI.

With an appropriate update set of nodes, $\mathcal{U}$, we can then update the edge set to obtain $\hat{E}^{EM}_{\ell_1}$.  In the following subsections we go over obtaining the update set $\mathcal{U}$ and completing the $E$ and $M$ steps.

\subsection{Obtaining Update Set: $\mathcal{U}$}\label{EM:U}

The update set will be a union of {\it candidate nodes}, $\mathcal{C}$, and {\it participant nodes}, $\mathcal{P}$.  Candidate nodes are nodes that have potentially been misclassified, and participant nodes are nodes where their estimated neighborhood sets have been potentially affected by misclassification.

If $\hat\gamma_s$ is a misclassification estimation for each node, then for a given threshold $q$, a reasonable way to define candidate set is as $\mathcal{C} = \{s \in V : \hat\gamma_s > q\}$, although our method is not bound to any procedure on determining the candidate set.

To obtain the participant nodes, first consider the following example. Assume $(r, s) \in E$ and $(s, t) \in E$ but $(r, t) \not\in E$.  If there were no misclassification in our data then $x_r|x_s \perp\!\!\!\!\perp x_t|x_s$, but if $x_s$ is a candidate node with some non-zero probability for misclassification, then we have
\begin{align}
\begin{split}
P(x_r=1, x_t=1|\hat{x_s}=1) &= P(x_s = \tilde{x_s})P(x_r=1, x_t=1|x_s=1)  \\
&\qquad  +P(x_s\neq \hat{x_s})P(x_r=1, x_t=1|x_s = -1) \\
&= (1-\gamma_s)P(x_r=1|x_s=1)P(x_t=1|x_s=1)  \\
&\qquad  +\gamma_s(x_r =1|x_s=-1)P(x_t=1|x_r=-1) \\
&\neq P(x_s=1|\hat{x_r}=1)P(x_t=1|\hat{x_r}=1)
\end{split}
\end{align}

Thus nodes are no longer independent as long as $\theta^*_{rs} \neq \theta^*_{st}$, and in the fitted network the edge $(r, t)$ may appear.  On the other hand, if $x_r$ was a candidate node, then $P(x_t = 1 | x_s = 1, x_r = 1) = P(x_t=1|x_s=1)$.  That is to say that if a node's shortest path to a candidate node in the true network is greater than or equal to 2, then that node's neighbors will still be chosen independently from the misclassification.  This is not only a useful heuristic for choosing an update set, but will also be a useful property when calculating weights for the EM fit.

Taking this into account, we set the update set to be $\mathcal{U} = N(N(\mathcal{C}))$, the neighbors of neighbors of the candidate nodes.  From here we have the participant nodes as all nodes in $\mathcal{U}$ that are not in $\mathcal{C}$,  i.e. $\mathcal{P} = \mathcal{U} \setminus \mathcal{C}$.

Lastly, let $s$ be the number of disjoint subgraphs induced by $\mathcal{U}$ and let $c_{max}$ be the largest number of candidate nodes in a single subgraph.  The computational complexity of the method is $O(sn2^{c_{max}})$, which can computationally tractable even with up to 20 candidates node in a single subgraph.  For the rest of the document, we assume $s = 1$, but for $s > 1$ the $E$ and $M$ steps still hold where a loop is run over each disjoint subgraph.

\subsection{E Step}\label{EM:E}
For the $k^{th}$ step in the EM update, for node $r \in \mathcal{U}$, we \cb{take the expectation over the lantent variabes $x_{r}$ \strike{ are interested in the expectation over the candidate nodes as their true state is a latent variable}}. 
Define the following three sets of parameters
\begin{align*}\theta_{\mathcal{U}\setminus r} &= \{\theta_{sr} ; s \in \mathcal{U}\} \\
\theta^{(k)}_{V\setminus\mathcal{U}\setminus r} &= \{\theta^{(k)}_{sr} ; s \not\in\mathcal{U}\}\\
\tilde{\theta}_{\setminus r} &= \theta_{\mathcal{U}\setminus r} \cup \theta^{(k)}_{V\setminus\mathcal{U}\setminus r} \end{align*} $\theta_{\mathcal{U}\setminus r}$ corresponds to the the neighborhood parameters for node $r$ that will be updated.  \cb{For} $s \not\in \mathcal{U}$, \cb{the corresponding edge parameter} $\theta^{(k)}_{sr}$ \cb{will not be updated \strike{is a neighbor of $r$ that will not be updated}}, and thus when running this update, the value $2\theta^{(k)}_{sr}x_rx_s$ is included as an offset in the logistic regression to account for their neighborhood effect.

We are interested in the penalized log likelihood \begin{align}\label{penEM}L_{\lambda}(\theta_{\mathcal{U}\setminus r}|\theta^{(k)}_{V\setminus \mathcal{U}\setminus r}, \tilde{\mathfrak{X}}) &= \tilde\ell_r(\theta_{\mathcal{U} \setminus r}; \theta^{(k)}_{V\setminus \mathcal{U} \cup r}, \tilde{\mathfrak{X}}) - \lambda \|\tilde{\theta}_{\setminus r}\|_1 \\&= \frac{1}{n}\sum_{i=1}^n\log P_{\tilde{\theta}_{\setminus r}}(\tilde{x}_r^{(i)}|\tilde{x}_{\mathcal{U} \setminus r}^{(i)}) - \lambda \|\tilde{\theta}_{\setminus r}\|_1  \end{align}

By including the offset terms in the regularization term, we ensure that the log likelihood will increase over a fixed parameter $\lambda$ \cb{\strike{as shown in Appendix}} \ref{apEM}.  Let $\Omega_{\mathcal{C}} = \{-1, +1\}^{|\mathcal{C}|}$, and for $z_c \in \Omega_{\mathcal{C}}$, let $\tilde{x}^{(i)}(z_c)$ be original observation with candidate nodes replaced by $z_c$.  An estimate of the expectation of this log likelihood is
\begin{align}
\hat{Q}_r(\theta_{\mathcal{U}\setminus r}|\theta^{(k)}, \hat\theta_{\setminus r}, \mathfrak{X}) &= \hat{E}_{\tilde{X}_{\mathcal{C}}|\tilde{X}^{(i)}_{V\setminus \mathcal{C}}; \theta^{(k)}}\left(\tilde\ell_r(\theta_{\mathcal{U} \setminus r}; \hat\theta_{V\setminus \mathcal{U} \cup r}, \tilde{\mathfrak{X}})\right) -\lambda \|\tilde{\theta}_{\setminus r}\|_1\\
&= \frac{1}{n} \sum_{i=1}^n\sum_{z_c \in \Omega_{\mathcal{C}}} \left[P_{\theta^{(k)}}(X_{\mathcal{C}} = z_c | \tilde{X}_{\mathcal{U}} = \tilde{x}^{(i)}_{\mathcal{U}}) \log P_{\tilde{\theta}_{\setminus r}}(\tilde{x}_r^{(i)}, z_c|\tilde{x}^{(i)}_{\mathcal{U}\setminus r})\right] \nonumber\\ & \quad \quad - \lambda \|\tilde{\theta}_{\setminus r}\|_1 \label{Estep1}
\end{align}

However, the joint probability $P_{\tilde{\theta}_{\setminus r}}(\tilde{x}_r^{(i)}|\tilde{x}^{(i)}_{\mathcal{U}\setminus r},z_c)$ is computationally intractable to maximize over unless $|\mathcal{C}|$ is very small.  We instead look only at conditional distributions, and consider the following estimate of the expectation
\begin{align}
\tilde{Q}_r(\theta_{\mathcal{U}\setminus r}|\theta^{(k)}, \hat\theta_{\setminus r}, \mathfrak{X}) &= \frac{1}{n} \sum_{i=1}^n\sum_{z_c \in \Omega_{\mathcal{C}}} \left[P_{\theta^{(k)}}(X_{\mathcal{C}} = z_c | \tilde{X}_{\mathcal{U}} = \tilde{x}^{(i)}_{\mathcal{U}}) \log P_{\tilde{\theta}_{\setminus r}}(\tilde{x}_r^{(i)}(z_c)|\tilde{x}^{(i)}_{\mathcal{U}\setminus r}(z_c))\right] \nonumber\\ & \quad \quad - \lambda \|\tilde{\theta}_{\setminus r}\|_1 \label{Estep2}
\end{align}

In Appendix \ref{apEM} it is shown for any set of observations $\tilde{X}$ and for any initial fit $\hat\theta$, there exists an open set of misclassification probabilities such that maximizing $\tilde{Q}_r$ will still result in an increase in the penalized likelihood $L_\lambda(\theta_{\mathcal{U}\setminus r}|\hat\theta_{V\setminus\mathcal{U}\setminus r})$.

The function $\tilde{Q}_r$, corresponds to a $\ell_1$-regularized weighted logistic regression. Each $P_{\theta^{(k)}}(\tilde{X}_{\mathcal{C}} = z_c | \tilde{X}_{\mathcal{P}} = \tilde{x}^{(i)}_{\mathcal{U}})$ can be calculated utilizing factorizations of Ising distribution where the partition function is cancelled out due to conditioning the probability.  A derivation of these probabilities is located in Appendix \ref{apWE}.

\subsection{M Step}\label{EM:M}

Noting that $\tilde{Q}_r$ corresponds to a weighted penalized logistic regression with an offset, we complete the M step maximization using the \texttt{glmnet} package in \texttt{R} \citep{friedman2009glmnet}.  We obtain the updated edge parameter estimates as
\begin{equation}
\theta^{(k+1)}_{\setminus r} = \left( \arg\min_{\theta_{\mathcal{U}\setminus r}\in\mathbb{R}^{|\mathcal{U}|-1}}\hat{Q}_r(\theta_{\mathcal{U}\setminus r}|\theta^{(k)}, \hat{\theta}_{\setminus r}, \tilde{\mathfrak{X}})\right) \cup \hat\theta_{V\setminus \mathcal{U} \cup r}
\end{equation}

With the updated edge set as
\begin{equation}
\hat{E}_{EM}^{(k+1)} = \{(s, t); \text{ if } (\theta^{(k+1)}_{\setminus s})_t \neq 0 \text{ and } (\theta^{(k+1)}_{\setminus t})_s \neq 0\}
\end{equation}

We show through simulations that this methodology tends to increase model selection performance of the underlying graphical model.

\section{Simulations}\label{sims}

The EM method uses information about the misclassification, and also leverages dependency/structure information which we have access to from the original fit as made formal in Section \ref{istheory:4}.

In the following simulation we demonstrate that candidate nodes will gain spurious connections due to misclassification, which can be overcome using the EM update.

One can also note that given misclassification information, a ``prior" weight based solely on misclassification information (i.e. agnostic of any structural dependency information) can be calculated as \begin{equation}\label{nullweight}P(\tilde{X}_{\mathcal{C}} = z_c) = \prod_{s \in \mathcal{C}} \gamma_s\end{equation} The EM method updates these state probabilities given dependency information.

\subsection{Simulation Parameters and Network Specification}\label{sims:spec}
\begin{figure}
\centering
\includegraphics[width=4in]{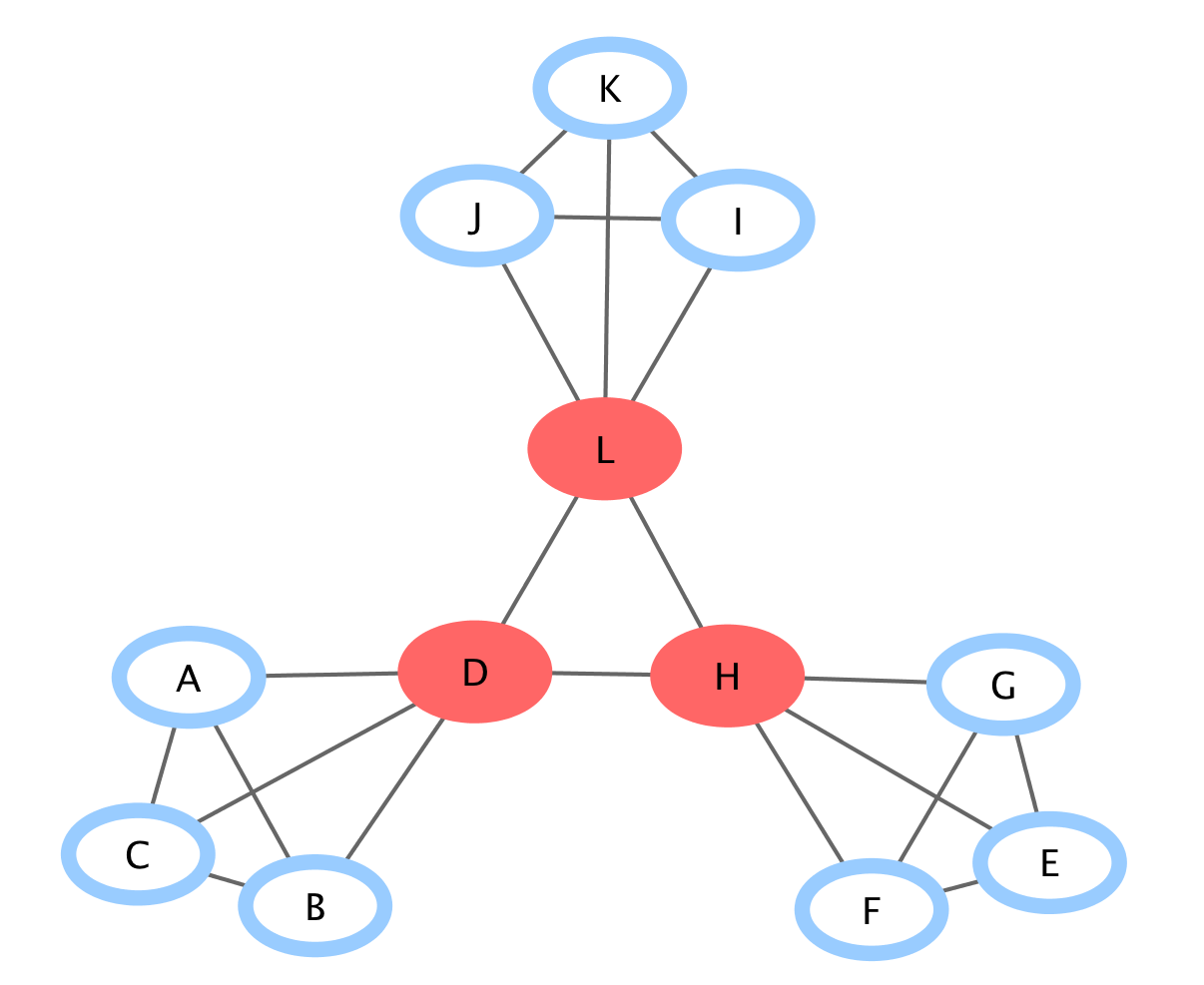}

\caption{The simulated network topology.  Red nodes (L, D, H) are {\it candidate nodes} which have a true misclassification probability of 60\% in half of observations.  Blue nodes correspond to participant nodes.  All non-zero edges have an equal weight = 1/2.}
\label{fig1}
\end{figure}

We ran the method on \gjh{a network of} 12 nodes ($p = 12$); Figure \ref{fig1} shows the topological structure of the network \gjh{over which} we \gjh{simulate}.  The intuition for this network topology is that the blue participants nodes will \cb{inform \strike{be informing}} the \cb{red} candidate \cb{\strike{red}} nodes.

The nodes $L, D, H$ are each potentially misclassified in 50\% of observations, where the probability of misclassification in these observations is 60\%.   We ran 1000 simulations with $n = 60$, and true edge parameters $\theta^*_{st} = \frac{1}{2}$ for $(s, t) \in E$. All Ising observations were simulated using the \texttt{IsingSampler} package in \texttt{R} \citep{epskamp2014isingsampler}.

\cb{Although nodes {\it L, D, H} are only misclassified in half of observations, the distribution unconditional on knowledge of the misclassification process is still a {\it Misclassified Ising Distribution} with non-zero misclassification parameters equal to $\gamma_{L} = \gamma_{D} = \gamma_{H} = 0.8$. }

\subsection{Fitted Models}\label{sims:fit}
The models we fit are

\begin{itemize}
\item[1.] {\it RWL} - minimizing (\ref{neilas})
\item[2.] {\it RWL Weighted} - minimizing (\ref{neilas}) with a weighted logistic regression using weights defined in (\ref{nullweight})
\item[3.] {\it RWL + EM} - Running an EM update for edges selected in RWL
\item[4.] {\it Weighted + EM} - Running an EM update for edges selected in RWL weighted
\end{itemize}

For the initial RWL and RWL Weighted fits, a range of tuning parameters were selected to obtain an ROC curve for candidate and participant nodes.  For the EM fits, the selected dependency was based off of the tuning parameter that maximized $P(True \,\, Positive) + (1-P(False\,\, Positive))$, and then a range of tuning parameters were simulated over to analyze the EM fits.

The first set of simulations look at only one EM update on our fit.  We then investigate the effect of further EM analyses.  We look at $RWL + 2EM$ and $RWL + 3EM$, which corresponds to running a second and third EM update to the on the $RWL$ fitted edge set.

\subsection{Results}\label{sims:res}
\begin{figure}
\centering
\includegraphics[width=6in]{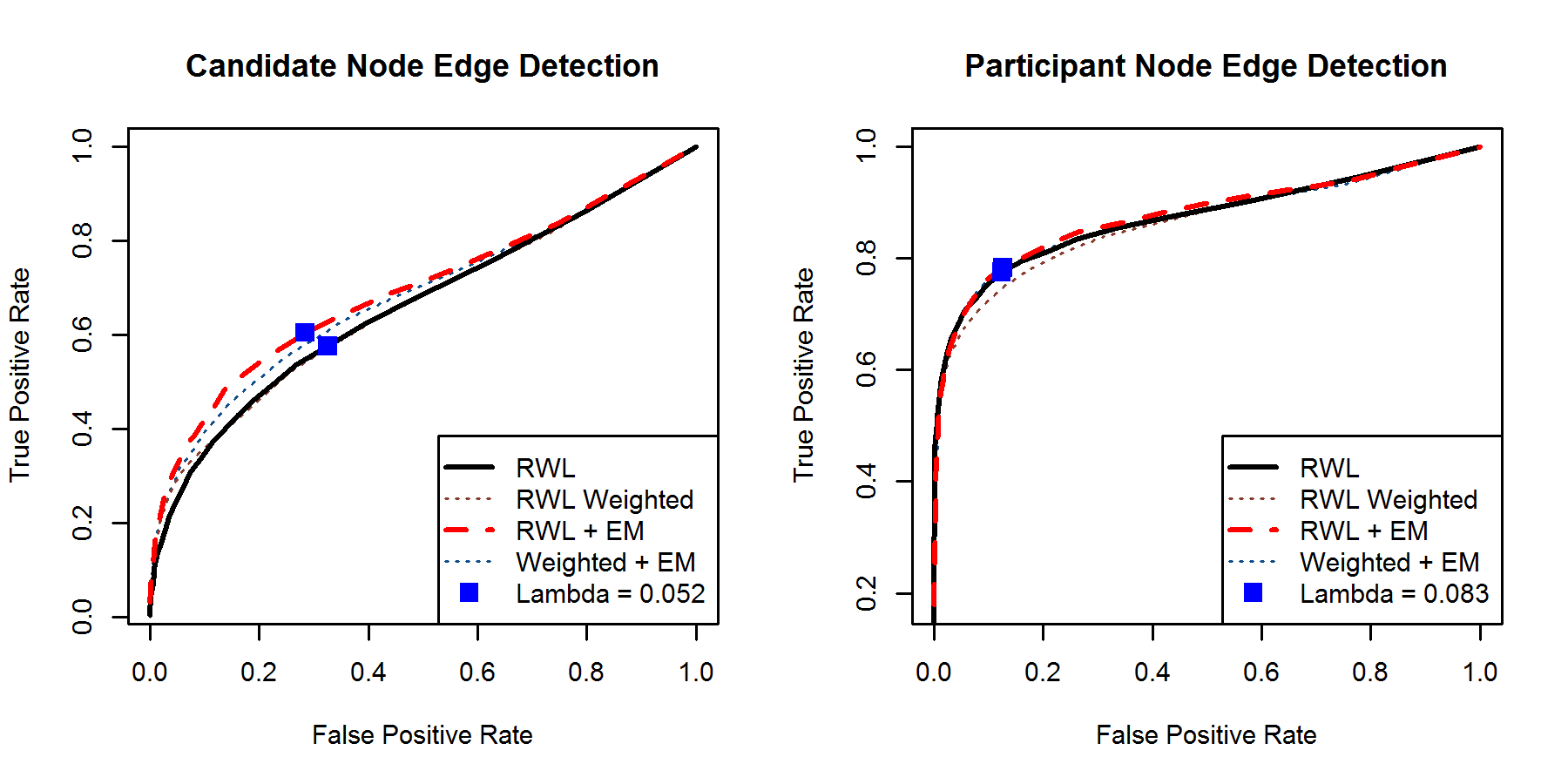}

\caption{Output from 1000 simulations with $n = 60$ and $p = 12$ showing True Positive vs True Negative and False Positive vs True Positive relationships for candidate and participant nodes.  Due to the symmetry in the topology of the graph, candidate and participant node results are aggregated.  }
\label{fig2}
\end{figure}

In Figure \ref{fig2} the RWL + EM fit performs at least as well or better than any other method.  Even when not changing the tuning parameter, an increase in classification performance is always observed. Specifically the AUC for candidate nodes increases from 0.6608 to 0.6945, and for participant nodes the AUC increases from 0.8729 to 0.8770.

Interestingly, basing the initial fit off of RWL seems to perform better than the weighted regularized logistic regression ({\it RWL Weighted}).  This is consistent with the proof given in Appendix \ref{apEM}, as the misclassification probability for a candidate node will be at most $P(X_r = \tilde{X}_r) = 0.5$ for {\it RWL Weighted}, and therefore this misclassification scenario is far from the open set $\mathbf{\Gamma}$ defined in Appendix \ref{apEM}.  The implication of this result is that misclassification information alone is not enough to provide a gain in model selection performance; dependency information must also be leveraged.

As shown in Section \ref{istheory:4} some dependency information is obtained in the $RWL$ fit, from which we have that $P(X_r = \tilde{X}_r|RWL) \approx 0$ for multiple observations, and therefore the {\it Regularized EM Theorem} in Appendix \ref{apEM} applies. Figure \ref{fig2} demonstrates this theorized increased in performance, and, as shown in Appendix \ref{apEM}, the increase will occur without needing to change the tuning parameter.

\begin{figure}
\centering
\includegraphics[width=6in]{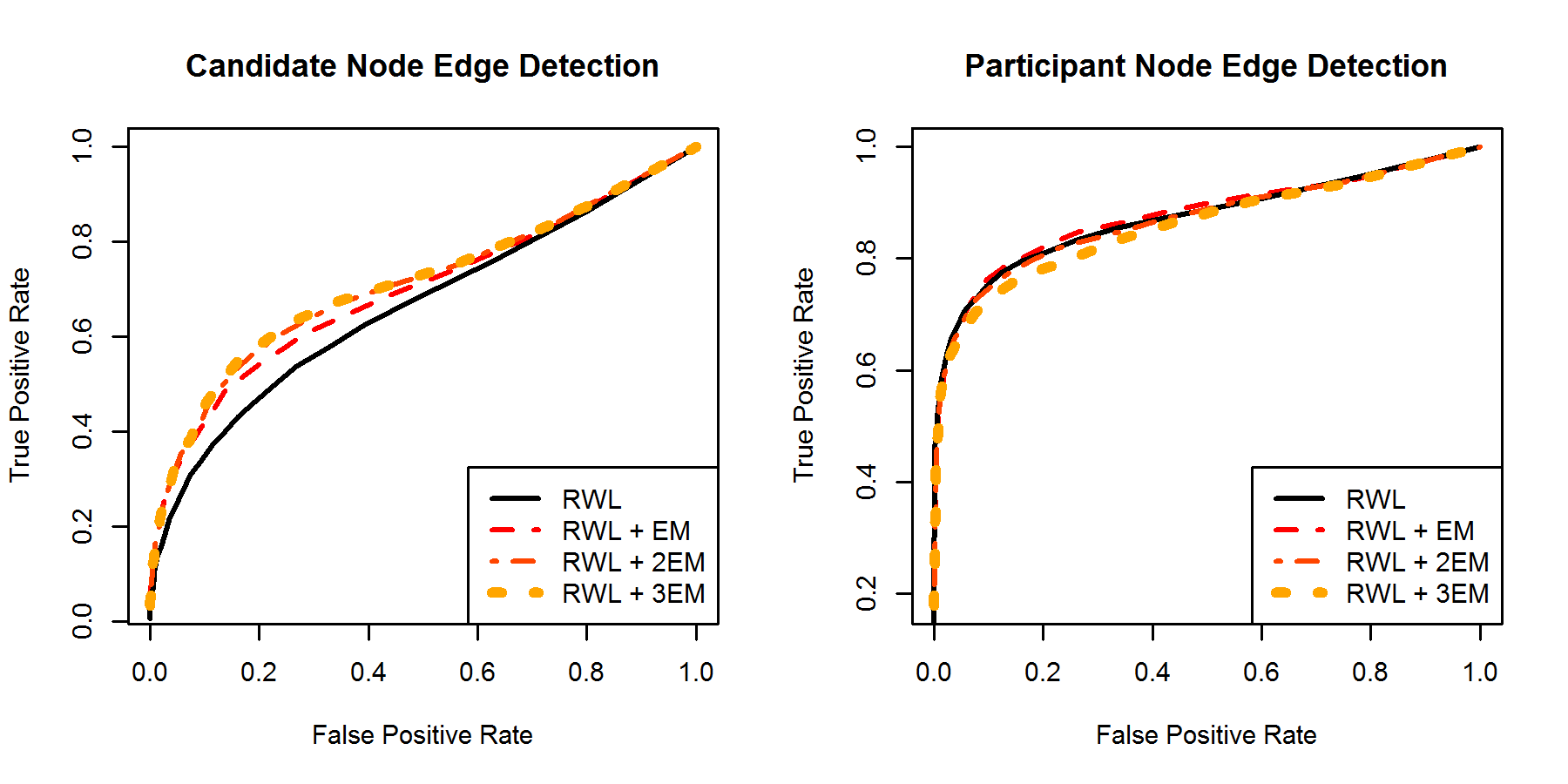}

\caption{Output from 1000 simulations with $n = 60$ and $p = 12$ showing True Positive vs True Negative and False Positive vs True Positive relationships for candidate and participant nodes for running multiple EM updates.  Note the decrease in performance for participant nodes for the 3rd EM update. }
\label{multEM}
\end{figure}

Figure \ref{multEM} shows the simulations results for running the EM update multiple times. Note that between EM updates  it is unlikely the probability that a node is in a given state will change drastically, therefore the {\it Regularized EM Theorem} does not apply.  This can be seen in Figure \ref{multEM}, as by the third EM update, there is a small decrease in participant node detection.  After the first EM update the participant node AUC is 0.8770, and it decreases to 0.8593 by the third update.

\section{fMRI Data Example Simulations}\label{fmri}
\citet{sinclair2017} documents a method for fitting an Ising model on task-fMRI data.  Each node in the graph corresponds to a specialized region of the cortex, and the classification is a discretization of a fit parameter corresponding to blood flow.  If the blood flow is above a certain threshold, the area of the cortex is considered active during the task.  Due to the inherent noise in the data, misclassification is certainly present.

\begin{figure}
\centering
\includegraphics[width=6in]{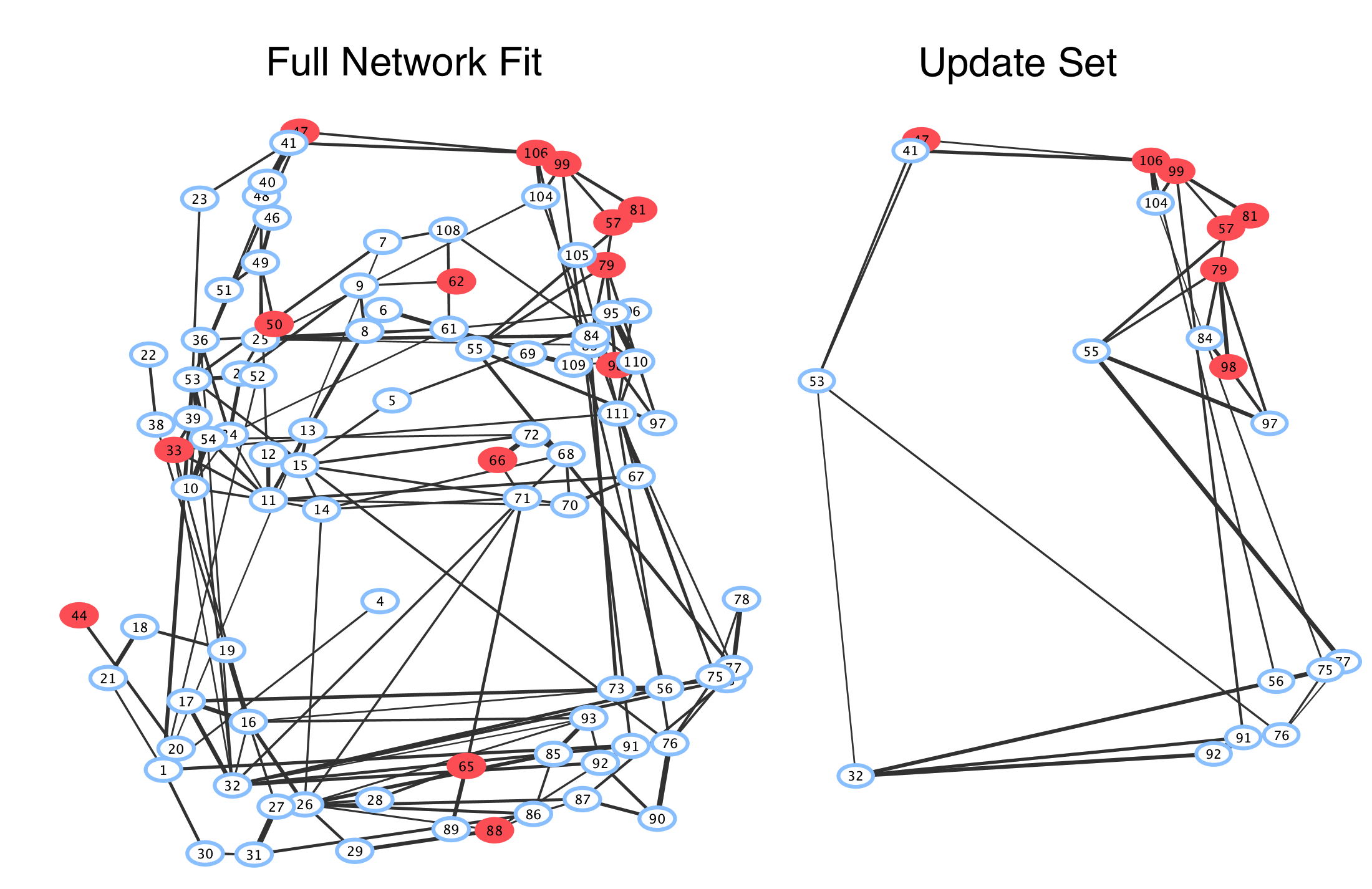}

\caption{The fitted connectome network from \citet{sinclair2017} and the update set $\mathcal{U}$.  Nodes are arrange in a superior (top-down) view of the cortex, where red nodes correspond to candidate nodes and blue nodes are all remaining nodes.}
\label{fig3}
\end{figure}

Figure \ref{fig3} shows the fit example from \citet{sinclair2017}, \gjh{using data} from the Human Connectome project \citep{van2013wu}, and the nodes were obtained via the parcellation documented in \citet{gordon2016generation}.  An estimate of the node's state was obtained \cb{\strike{from the method's }}\gjh{\strike{???? which method?}} \cb{\strike{classification procedure} by investigating the p-values used for the classification procedure. \strike{, and it was found that 14 out of 111 regions had a greater than 90\% chance of misclassification over 12\% of the time}  14 out of the 111 regions were found to be closer to the p-value threshold more often, being within 5\% of the p-value threshold at over 12\% of the time.}  In Figure \ref{fig3}, these regions are colored in red.

\subsection{Choosing Update Set $\mathcal{U}$}\label{fmri:U}

A useful consequence of the network fit we have, is that the update set as defined in Section \ref{EM:U} is a disjoint union of $s = 4$ disjoint subgraphs.  Therefore, we run our simulations on the largest of the subgraphs denoted as the update set in Figure \ref{fig3}.
This \gjh{ corresponds} to our $p = 20$ node network topology that we \gjh{use} for simulations.

\subsection{Simulation Parameters}\label{fmri:sim}

We ran 500 simulations with $n = 200$, corresponding to the size of the original dataset.  Edge parameters in the simulation were selected to correspond to edge parameters from the original fit, however non-zero edges  were smoothed towards the average of all edge parameters.

Participant nodes were then misclassified in 50\% of observations with a misclassification probability of 75\%.  Thus, the overall misclassification rate is similar to the observed dataset.

Based off of the results from Section \ref{sims}, we only compare the RWL + EM and RWL, where a range of tuning parameters is selected for each method.

\subsection{Results}\label{fmri:sim}

\begin{figure}
\centering
\includegraphics[width=6.5in]{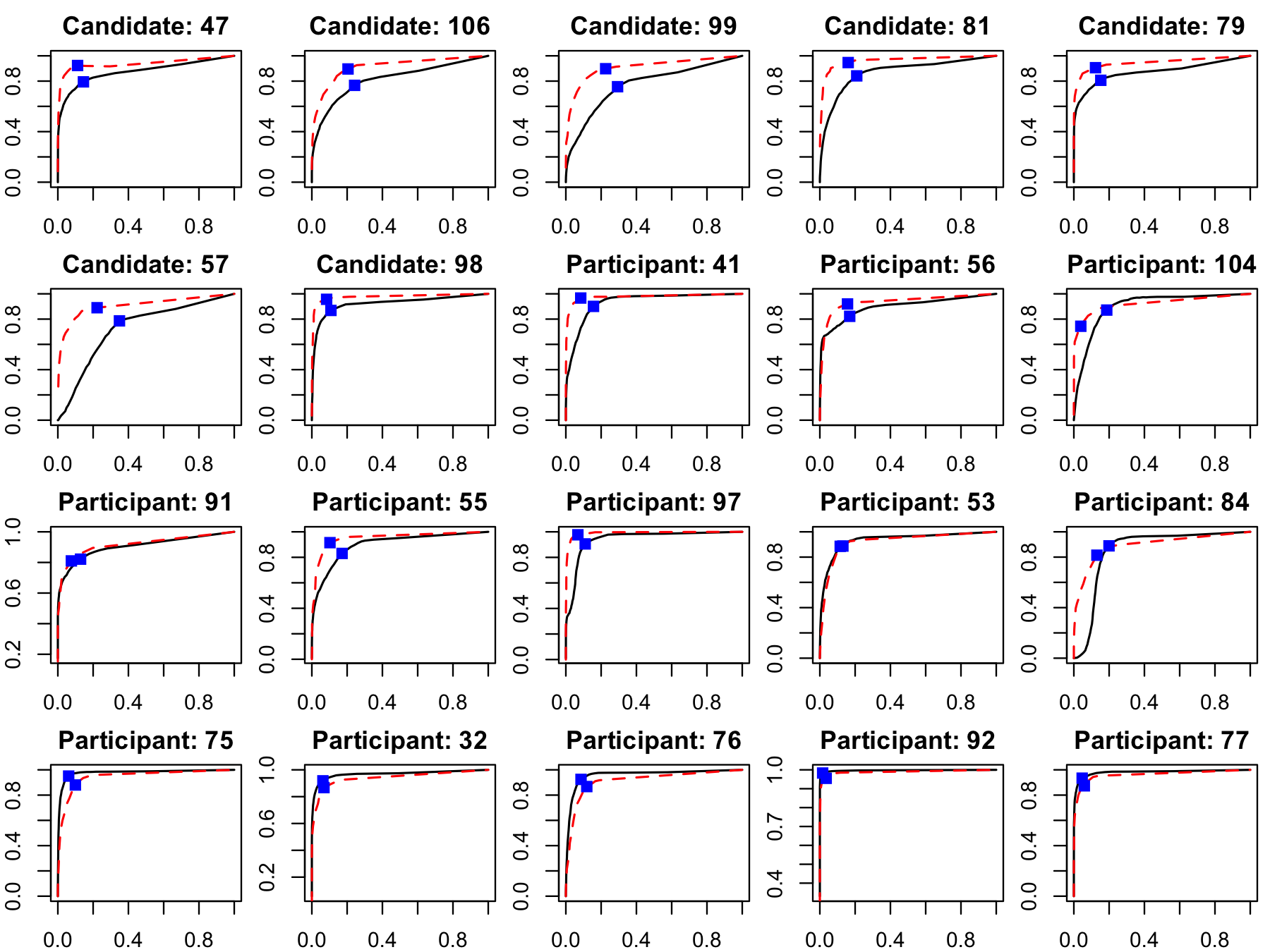}

\caption{True Positive (x-axis) vs False Positive (y-axis) rate per node.  Red corresponds to the EM updated curve, and black corresponds to the original Ravi fit.  The optimal tuning parameter is labelled with blue square on the RWL fit line, and the corresponding tuning parameter is labelled on the EM fit line. }
\label{fig4}
\end{figure}

Figure \ref{fig4} shows the True Positive vs False Positive relationship.  A consistent increase in classification performance is observed for the first 13 nodes.  The overall error rate decreases for the neighborhood of candidate nodes drops from from 21.1\% to 10.0\% when choosing the optimal tuning parameter for the EM fit.  If the tuning parameter is not changed for the EM fit, we still see an decrease in the error from 21.1\% to 14.6\%.  There does appear to be a small decrease in performance for participant nodes that were not a direct neighbor with a candidate however this difference contributed to less than a 3\% increase in false positives and false negatives.

\begin{figure}
\centering
\includegraphics[width=4in]{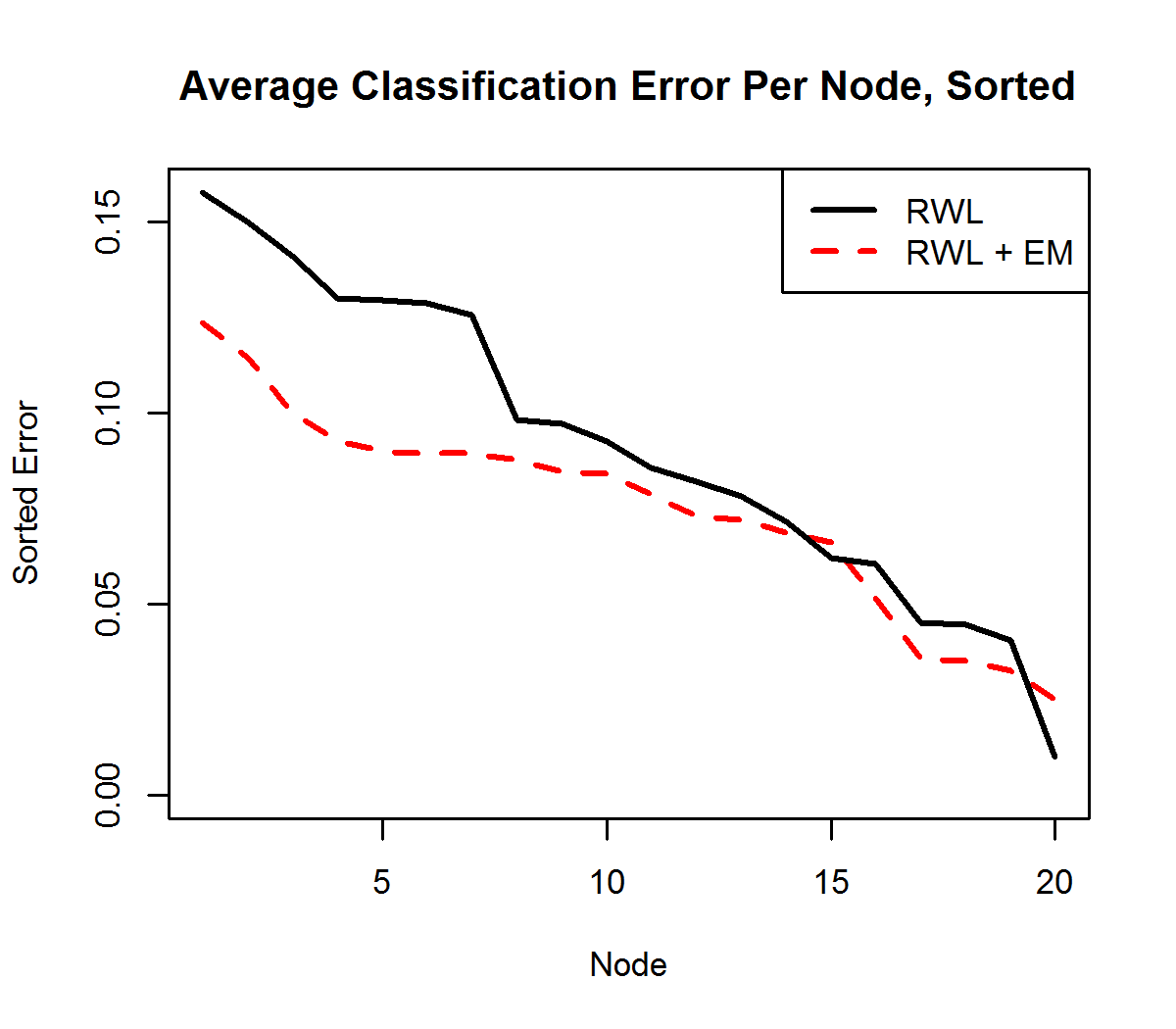}

\caption{Error rates for each node, sorted within method.  Error rates are average number of False Positives and False Negatives per node}
\label{fig5}
\end{figure}

Figure \ref{fig5} orders the nodes by overall error rate across simulations for the two different methods.  The decrease in error rate is consistently better after running the EM fit.

\begin{figure}
\centering
\includegraphics[width=6.5in]{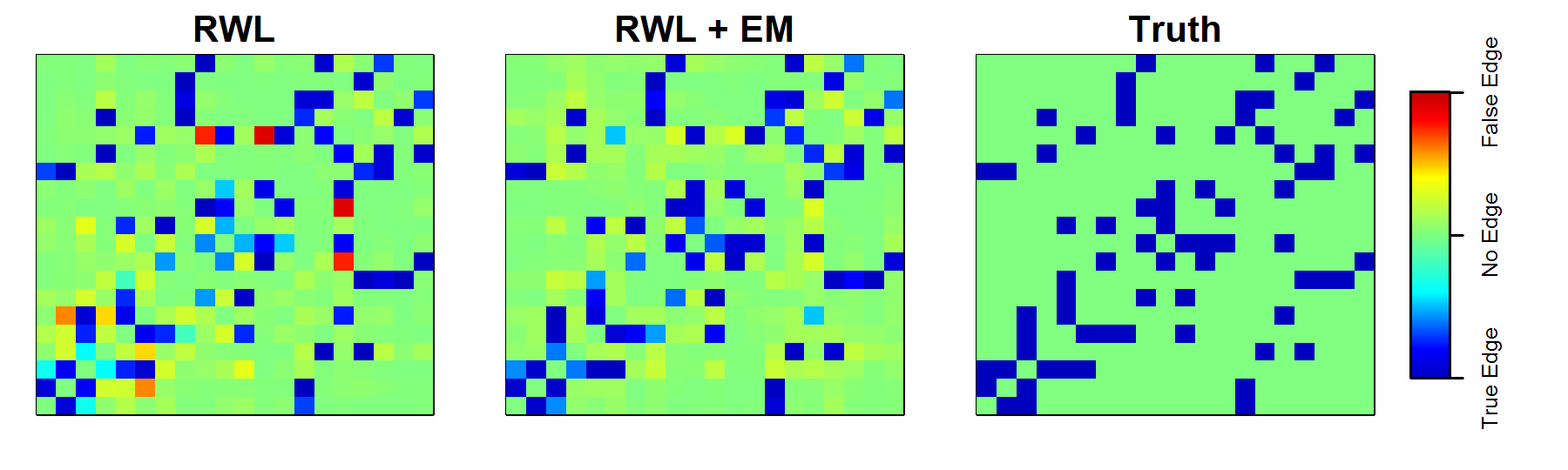}

\caption{Overall adjacency matrix selection.  True edges range from green to blue, where a darker blue corresponds to more true positives.  False edges range from red to green, where a darker red corresponds to more false positives.}
\label{fig6}
\end{figure}

\begin{figure}
\centering
\includegraphics[width=6in]{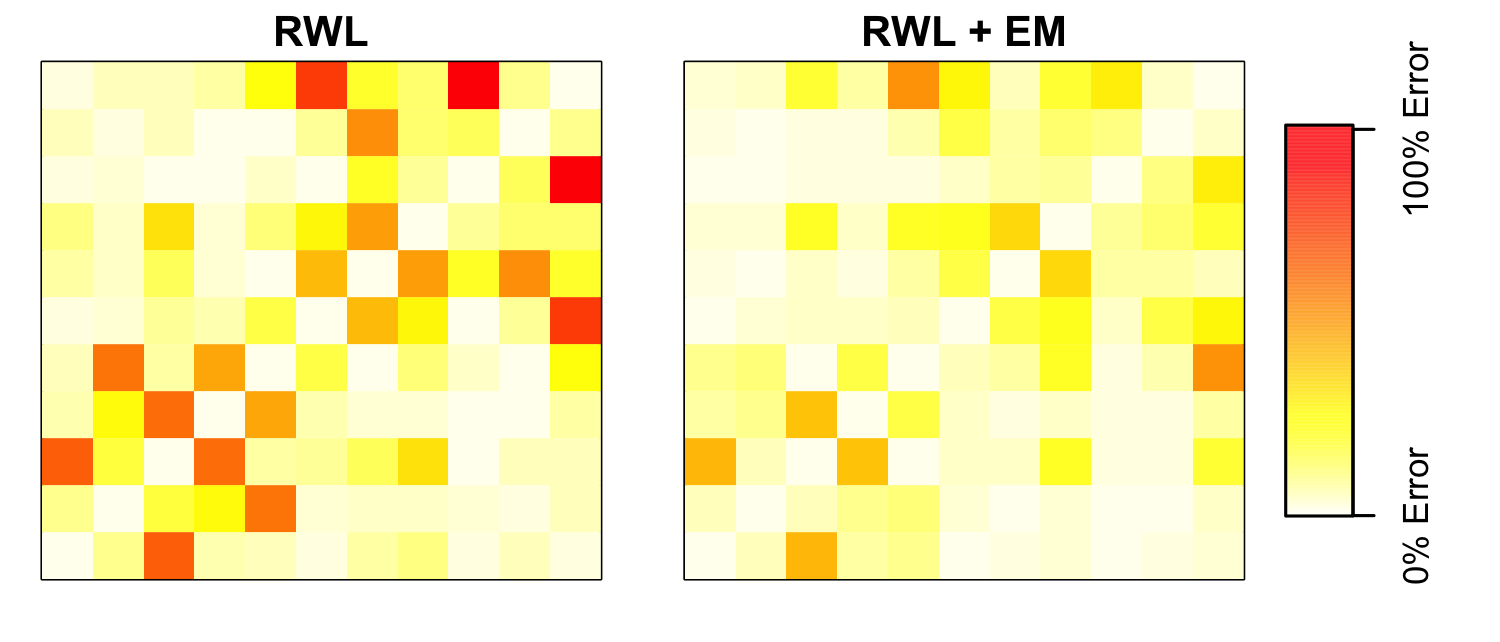}

\caption{Error rates per edge \cb{for nodes that are candidate nodes or direct neighbors of candidate nodes.\strike{,where} The} error rate is calculated as False Negatives + False Positives. Darker red corresponds to a higher error rate. }
\label{fig7}
\end{figure}

Figure \ref{fig6} plots the adjacency matrix for $\mathcal{U}$.  This plot has a few interesting characteristics.  The red areas, which correspond to false edges that were selected often for the RWL fit tend to correspond to edges between participant nodes that are highly connected to candidate nodes.  The error rate is particularly high for nodes 104, 84, and 55.  Figure \ref{fig7} looks only at error rate, and focusses on nodes that had at least one neighbor with a candidate node.

\section{Conclusion}
In this paper we introduce the misclassified Ising model.  We show that under suitable misclassification assumptions RWL can still be used as a model selection technique.  We then show that RWL can be extended in order to account for misclassification.  Sections \ref{sims} \cb{and} \gjh{\strike{and Section}} \ref{fmri} show simulation results for a symmetric network and for a network obtained from fMRI data.

The fMRI node states correspond to discretizations of a continuous variable and therefore provide a useful setting for discussing misclassification. Depending on the discretization method used to determine the latent state, acquiring an estimate for the probability of misclassification is potentially straightforward.

In both cases, the EM-based algorithm is shown to provide significant performance gains in model selection.  Given a binary network data set with an estimated misclassification probability, one can therefore obtain more reliable connections between nodes within the update set $\mathcal{U}$ by performing this update.

The method is \gjh{computationally} constrained by the greatest number of candidate nodes within  the largest disjoint subnetwork of the update set $\mathcal{U}$.  However, this computational complexity depends only linearly on the number of remaining nodes in the update set.  Therefore even with a high degree dataset, if there are few candidate nodes, this method can still be tractable.

The analysis in this paper can be extended easily to the signed edge selection as discussed in \citet{ravikumar2010high}.  The EM approach can also be extended to the Potts model corresponding to multiple states per node, although this would serve to further increase the computational complexity. Future work within the misclassified Ising framework could be to understand the effect of dependent misclassification across nodes on the misclassified score and information functions.


\bibliographystyle{chicago}
\bibliography{embibJASA}

\begin{appendices}
\section{Proof of Extended Theorem 1}
In this appendix we state the assumptions for extended theorem 1 and complete the proof.

\subsection{Assumptions}

In order to prove {\it extended theorem 1} we need to make assumptions $\tilde{A1}$, $\tilde{A2}$, $\tilde{A3}$.  Assumptions $\tilde{A1}$ and $\tilde{A2}$ are analogous to \citet{ravikumar2010high} except under the misclassified information matrix.  Assumption $\tilde{A3}$ bounds the amount of misclassification in our data.

Define $S = \{(r, t) \in V \times V | t\in \mathcal{N}(r)\}$.

Assume the following assumptions hold uniformly for all $r \in V$:

{\it ($\tilde{A1}$) Dependency Condition}. For the misclassified information matrix and for the sample covariance matrix, there exists a constants $C_{min}, D_{max} > 0$ such that \begin{align} \Lambda_{min}((\tilde{Q}^*_r)_{SS}) &\geq C_{min}\\
\Lambda_{max}(E_{\gamma, \theta^*}[X_{\setminus r}X_{\setminus r}^T]) &\leq D_{max}
\end{align}

{\it ($\tilde{A2}$) Incoherence Condition}. There exists $\alpha \in (0, 1]$ such that
\begin{equation}
\|\tilde{Q}^*_{S^cS}(\tilde{Q}^*_{SS})^{-1}\|_\infty \leq 1-\alpha
\end{equation}

{\it ($\tilde{A3}$) Misclassification Condition}. For $C_{min}$ and $D_{max}$ as defined in ($\tilde{A1}$), and $\alpha$ as defined in ($\tilde{A2}$), we assume
\begin{equation}
S_{max} \leq \frac{C^2_{min}\alpha^2}{400D_{max}d(2-\alpha)^2}
\end{equation}

\subsection{Proof}
Within this proof we drop the node-specific subscript $r$.  The proof is done within node, and a union bound is applied to obtain the result across nodes.

Define the sample misclassified information as
\begin{equation} \tilde{Q}^n = -\hat{E}(\nabla W^n(\theta^*)) \end{equation}
In \citet{ravikumar2010high}, Lemma 5, 6, and 7 can be applied to show that if $\tilde{\mathfrak{X}}$ is such that $\tilde{A1}$ and $\tilde{A2}$ hold for $\tilde{Q}^n$, then the assumptions will hold for with high probability for $\tilde{Q}^*$ for $n = \Omega(d^3 \log p)$.  These lemmas directly apply to the misclassified case since their only dependence on the Ising distribution is that $\tilde{Q}^n- \tilde{Q}^*$ can be written as an iid mean of bounded observations, which still holds.

Therefore, to complete the proof it suffices to show that {\it Extended Theorem 1} is true only for observations where the event $M = \{\tilde{\mathfrak{X}} : \tilde{A1} \text{ and } \tilde{A2} \text{ hold for }\tilde{Q}^n\}$ occurs.  This corresponds to Proposition 1 of \citet{ravikumar2010high}.

Define $\tilde{\lambda_n} = \lambda_n - \frac{4(2-\alpha)}{\alpha}S_{max}$.  We can use Lemma 3, and Lemma 4 from \citet{ravikumar2010high} to show {\it Extended Theorem 1} holds when $M$ occurs.  In order to utilize these lemmas we need to establish an upper bound for the misclassified score function with high probability, and we need to establish an upper bound for the quantity $\lambda_nd$.  The following lemma proven in Appendix \ref{lemma}. established an upper bound on the misclassified score function.

{\bf Lemma}.  {\it For the specified incoherence parameter $\alpha \in (0,1]$, we have }
\begin{equation} P\left(\|W^n\|_\infty \geq \frac{\lambda_n}{4}\right) = O(\exp(-K\tilde\lambda^2_nn)) \end{equation} {\it for $K$ independent of $(n, d, p)$ and for $\lambda_n \geq \frac{16(2-\alpha)}{\alpha}\left(\sqrt{\frac{\log p}{n}} + \frac{S_{max}}{4}\right)$}

In order to establish bounds for $\lambda_nd$, set $n > \frac{400^2D^2_{max}}{C^4_{min}}\frac{(2-\alpha)^4}{\alpha^4}d^2\log p$, then by applying assumption ($\tilde{A3}$) on $S_{max}$, and since $\frac{\alpha}{2-\alpha} \leq 1$ we have
\begin{align}
\begin{split}
\lambda_nd &=  \frac{16(2-\alpha)}{\alpha}\left(\sqrt{\frac{\log p}{n}} + \frac{S_{max}}{4}\right)d \\
&< \frac{32C^2_{min}\alpha}{400D_{max}(2-\alpha)}\\
&< \frac{C^2_{min}}{10D_{max}}
\end{split}
\end{align}

With these technical results we can complete the proof of {\it extended theorem 1} as presented in \citet{ravikumar2010high}.

\subsubsection{Proof of Lemma} \label{lemma}
Let $W^n_u$ be the $u^{th}$ component of $W^n$.  Note that $W^n_u$ is the iid mean of $n$ random variables that are bounded between [-2,2].  Therefore by Azuma-Hoeffding inequality \citep{hoeffding1963probability}, we have
\begin{equation}\label{l1eq}
P(|W^n_u - E(W^n_u)| > \delta) \leq 2 \exp\left(-\frac{n\delta^2}{8}\right)
\end{equation}
for any $\delta > 0$.  Note that for any $x, y, z \in \mathbb{R}$ we have, $|x| > |z| + |y| \Rightarrow |x - y| > |z|$.  Applying this to (\ref{l1eq}) gives
\begin{equation}\label{l1eq2}
P(|W^n_u| > \delta + |E(W^n_u)|) \leq P(|W^n_u - E(W^n_u)| > \delta) \leq 2 \exp\left(-\frac{n\delta^2}{8}\right)
\end{equation}

We can bound (\ref{l1eq2}) from below by setting $\delta = \frac{\alpha \lambda_n}{4(2-\alpha)} - |E(W^n_u)|$, and noting that $\frac{\alpha}{2-\alpha} \leq 1$.  We get
\begin{equation}\label{l1eq3}P\left(|W^n_u| > \frac{\lambda_n}{4}\right) \leq P\left(|W^n_u| >  \frac{\alpha \lambda_n}{4(2-\alpha)} \right)\end{equation}

We bound (\ref{l1eq2}) from above as follows
\begin{align}\label{l1eq4}
\begin{split}
2\exp\left(-\frac{n\delta^2}{8}\right) &= 2\exp\left(-\frac{n}{8}\left[\frac{\alpha\lambda_n}{4(2-\alpha)}-|E(W^n_u)|\right]^2\right) \\
&\leq 2\exp\left(-\frac{n}{8}\left[\frac{\alpha\lambda_n}{4(2-\alpha)}-S_{max}\right]^2\right) \\
&= 2\exp\left(-\frac{n}{8}\left[\frac{\alpha\tilde\lambda_n}{4(2-\alpha)}\right]^2\right)
\end{split}
\end{align}

Combining (\ref{l1eq2}), (\ref{l1eq3}), (\ref{l1eq4}) finishes the proof of the lemma.

\section{Proof of Regularized EM Approach} \label{apEM}

In this appendix we show the following.

\noindent {\bf Regularized EM Theorem}. {\it For data $\tilde{\mathfrak{X}}$, for $\hat\theta$ the parameter estimate from the {\it RWL} fit, and for $\theta^*$ the parameter estimate from the first EM update, there exists an open set of misclassification laws $\mathbf{\Gamma}$ such that for the marginal penalized likelihood of our data as defined in Equation (\ref{penEM}) we have that }\begin{equation}L_\lambda(\theta^*_{\mathcal{U}\setminus r} | \hat\theta_{V\setminus \mathcal{U}\setminus r}, \tilde{\mathfrak{X}}) \geq L_\lambda(\hat\theta_{\mathcal{U}\setminus r} | \hat\theta_{V\setminus \mathcal{U}\setminus r}, \tilde{\mathfrak{X}})\end{equation}

For notational convenience, we suppress the parameters $\hat\theta_{V\setminus\mathcal{U}\setminus r}$, and we refer to our parameters of interested simply as $\theta$ as they do not change throughout the proof.

For $z_c$ as the latent states, by following the proof of the EM given in \citet{little2002statistical} we have the following relationship for the marginal likelihoods, which still holds when the regularization parameter is added
\begin{align}
L_\lambda(\theta | \tilde{\mathfrak{X}}) &= \sum_{i=1}^n\sum_{z_c \in \Omega_\mathcal{C}}P_{\hat\theta}(z_c | \tilde{X}^{(i)}) \log(P_\theta(\tilde{x}^{(i)}_r, z_c| \tilde{X}^{(i)}_{\setminus r})) \nonumber\\ &\quad - \sum_{i=1}^n\sum_{z_c \in \Omega_\mathcal{C}}P_{\hat\theta}(z_c | \tilde{X}^{(i)})\log(P_\theta(z_c|\tilde{X}^{(i)})) + \lambda\|\theta\|_1 \label{EMentropy}\\
&= A_\Gamma(\theta) + B_\Gamma(\theta)+ \lambda\|\theta\|_1\nonumber
\end{align}

Where $A_\Gamma(\theta)$ and $B_\Gamma(\theta)$ correspond the two large summations in equation (\ref{EMentropy}).  $\Gamma$ is included in the notation for these functions to emphasize their dependence on the misclassification scheme.

For $B_\Gamma(\theta)$ we have that by Gibb's inequality, $B_\Gamma(\theta) \geq B_\Gamma(\hat\theta)$ for all $\theta$, and for all $\Gamma$. Therefore $B_\Gamma(\theta)$ will increase at $\theta^*$.  Our goal is thus to show that $A(\theta) + \lambda\|\theta\|_1$ will increase.

Choose the misclassification setting $\Gamma'$ such that $\prod_{s \in \mathcal{C}}P(z_s \neq \tilde{x}^{(i)}_s) = 1$.  Define $z^{(i)}_{\Gamma'}$ component-wise as $(z^{(i)}_{\Gamma'})_s = -\tilde{x}^{(i)}_s$.  Under this $\Gamma'$, we have the following representation for  $A_{\Gamma'}(\theta)$
\begin{align}
A_{\Gamma'}(\theta) &=  \sum_{i=1}^n\sum_{z_c \in \Omega_\mathcal{C}}P_{\hat\theta}(z_c | \tilde{X}^{(i)}) \log(P_\theta(\tilde{x}^{(i)}_r(z_c)| \tilde{x}^{(i)}(z_c)_{\setminus r})P_\theta(z_c | \tilde{x}^{(i)}_{\setminus r})) \\
&= \sum_{i=1}^nP(z^{(i)}_{\Gamma'} | \tilde{x}^{(i)}) \log (P_\theta(\tilde{x}_r^{(i)}(z_{\Gamma'}^{(i)})| \tilde{x}^{(i)}(z^{(i)}_{\Gamma'})))
\end{align}

For this selection of $\Gamma'$ we have that $\theta^*$ is chosen to maximize $A_{\Gamma'}(\theta) + \lambda\|\theta\|_1$, and therefore $A_{\Gamma'}(\theta^*) + B_{\Gamma'}(\theta^*) + \lambda\|\theta^*\| \geq A_{\Gamma'}(\hat\theta) + B_{\Gamma'}(\hat\theta) + \lambda\|\hat\theta\|  $.  Since $A_\Gamma(\theta) + B_\Gamma(\theta) + \lambda\|\theta\|_1$ is continuous in $\Gamma$, there exists an open set $\mathbf{\Gamma}$ such that if $\Gamma \in \mathbf{\Gamma}$ then $L_\lambda(\theta^*_{\mathcal{U}\setminus r} | \hat\theta_{V\setminus \mathcal{U}\setminus r}, \tilde{\mathfrak{X}}) \geq L_\lambda(\hat\theta_{\mathcal{U}\setminus r} | \hat\theta_{V\setminus \mathcal{U}\setminus r}, \tilde{\mathfrak{X}})$ as needed.

\section{Calculating Weights for E-step}\label{apWE}
Here we calculate the weights $P_{\theta^{(k)}}(X_{\mathcal{C}} = z_c | \tilde{X}_{\mathcal{U}} = \tilde{x}^{(i)}_{\mathcal{U}})$ from equation (\ref{Estep1}).  In these calculations we assume we have $\gamma^i_s$ corresponding to the misclassification probability for node $s$ at observation $i$.

We remove the subscript for estimate $\theta^{(k)}$, and superscript for observation ${(i)}$ for notational convenience. Rearranging conditional and joint probabilities give us
\begin{align}
P(X_{\mathcal{C}} = z_c | \tilde{X}_{\mathcal{U}} = \tilde{x}_{\mathcal{U}}) &= \frac{P(X_{\mathcal{C}} = z_c, \tilde{X}_{\mathcal{C}} = \tilde{x}_{\mathcal{C}}, \tilde{X}_{\mathcal{P}} = \tilde{x}_{\mathcal{P}})}{P(\tilde{X}_{\mathcal{C}} = \tilde{x}_{\mathcal{C}}, \tilde{X}_{\mathcal{P}} = \tilde{x}_{\mathcal{P}})} \\
&= \frac{P(X_{\mathcal{C}} = z_c, \tilde{X}_{\mathcal{P}} = \tilde{x}_{\mathcal{P}})}{P(\tilde{X}_{\mathcal{C}} = \tilde{x}_{\mathcal{C}}, \tilde{X}_{\mathcal{P}} = \tilde{x}_{\mathcal{P}})} P(\tilde{X}_{\mathcal{C}} = \tilde{x}_{\mathcal{C}} | X_{\mathcal{C}} = z_c, \tilde{X}_{\mathcal{P}} = \tilde{x}_{\mathcal{P}}) \label{AC3}
\end{align}

The conditional probability in (\ref{AC3}) gives the proportion of the weight associated with the observed misclassification probability. This is calculated as

\begin{align}
P(\tilde{X}_\mathcal{C} = \tilde{x}_\mathcal{C} |  X_{\mathcal{C}} = z_c, \tilde{X}_{\mathcal{P}} = \tilde{x}_{\mathcal{P}}) &= P(\tilde{X}_\mathcal{C} = \tilde{x}_\mathcal{C} |  X_{\mathcal{C}} = z_c) \\
&= \prod_{s \in C}(\gamma_s\mathbf{1}(\tilde{x}_s \neq (z_c)_s) + (1-\gamma_s)\mathbf{1}(\tilde{x}_s = (z_c)_s)) \\
&= c(z_c, \tilde{x}_\mathcal{U})
\end{align}

The ratio of probabilities gives the weight of the observation associated with the estimated dependency structure.  Define $A(x_\mathcal{C}, x_\mathcal{P}) = \sum_{(s, t) \in E_{\mathcal{U}}} \theta^{(t)}_{st}x_sx_t$; this corresponds to the association between nodes in $\mathcal{U}$ as it relates to the full distribution given in (\ref{eq:1}).  From to the selection of $\mathcal{U}$ the ratio of probabilities factors allowing this calculation to ignore nodes outside of $\mathcal{U}$.

\begin{align}
\frac{P(X_{\mathcal{C}} = z_c, \tilde{X}_{\mathcal{P}} = \tilde{x}_{\mathcal{P}})}{P(\tilde{X}_{\mathcal{C}} = \tilde{x}_{\mathcal{C}}, \tilde{X}_{\mathcal{P}} = \tilde{x}_{\mathcal{P}})} &= \frac{B \exp(A(x_\mathcal{C}, x_\mathcal{P}))}{B \sum_{z_c' \in \Omega_c}c(z_c', \tilde{x}_\mathcal{U})} \\
&= \frac{\exp(A(x_\mathcal{C}, x_\mathcal{P}))}{\sum_{z_c' \in \Omega_c}c(z_c', \tilde{x}_\mathcal{U})}
\end{align}

Where $B$ in the above equation corresponds to the potential from all nodes outside of $\mathcal{U}$.
\end{appendices}

\end{document}